\documentclass[11pt,a4paper]{article}

\usepackage{subcaption} 
\usepackage[margin=1in]{geometry} 

\usepackage{jinstpub}
\usepackage[switch]{lineno}

\usepackage{lineno}

\title{Numerical and experimental investigations of a microwave interferometer for the negative ion source SPIDER}
\author[a,b,1]{R. Agnello\note{Corresponding author.},}
\author[b]{R.~Cavazzana}
\author[a]{I.~Furno}
\author[a]{R.~Jacquier}
\author[b]{R.~Pasqualotto}
\author[b]{E.~Sartori}
\author[b]{and G.~Serianni}
\affiliation[a]{\'{E}cole Polytechnique Fédérale de Lausanne (EPFL), Swiss Plasma Center (SPC),\\
CH-1015 Lausanne, Switzerland}
\affiliation[b]{Consorzio RFX,
Corso Stati Uniti 4, Padova, Italy}

\emailAdd{riccardo.agnello@epfl.ch}

\abstract{The electron density close to the extraction grids and the co-extracted electrons represent a crucial issue when operating negative ion sources for fusion reactors. An excessive electron density in the plasma expansion region can indeed inhibit the negative ion production and introduce potentially harmful electrons in the accelerator. Among the set of plasma and beam diagnostics proposed for SPIDER upgrade, a heterodyne microwave (mw) interferometer at 100 GHz is currently being explored as a possibility to measure electron density in the plasma extraction region. The major issue in applying this technique in SPIDER is the poor accessibility of the probing microwave beam through the source metal walls and the long distance of 4 m at which mw modules should be located outside the vacuum vessel. Numerical investigations in a full-scale geometry showed that the power transmitted through the plasma source apertures was above the signal-to-noise ratio threshold for the microwave module sensitivity. An experimental proof-of-principle of the setup to assess the possibility of signal phase detection was then performed. The microwave system was tested on an experimental full-scale test-bench mimicking SPIDER viewports accessibility constraints, including the presence of a SPIDER-like plasma. The outcome of first tests revealed that, despite the geometrical constraints, in certain conditions, the phase detection, and, therefore, electron density measurements are possible. The main issue arises from decoupling the one-pass signal from spurious multipaths generated by mw beam reflections, requiring signal cross correlation analysis. These preliminary tests demonstrate that despite the 4 m distance between the mw modules and the presence of metal walls, plasma density measurement is possible when the 80-mm diameter ports are available. In this contribution, we discuss the numerical simulations, the preliminary experimental tests and suggest design upgrades of the interferometric setup to enhance signal transmission.}
\keywords{mm-wave interferometry, plasma density, negative ion source}

\begin{document}
\maketitle
\flushbottom

\section{Introduction}
\begin{figure}
    \centering
    \includegraphics[width=\linewidth]{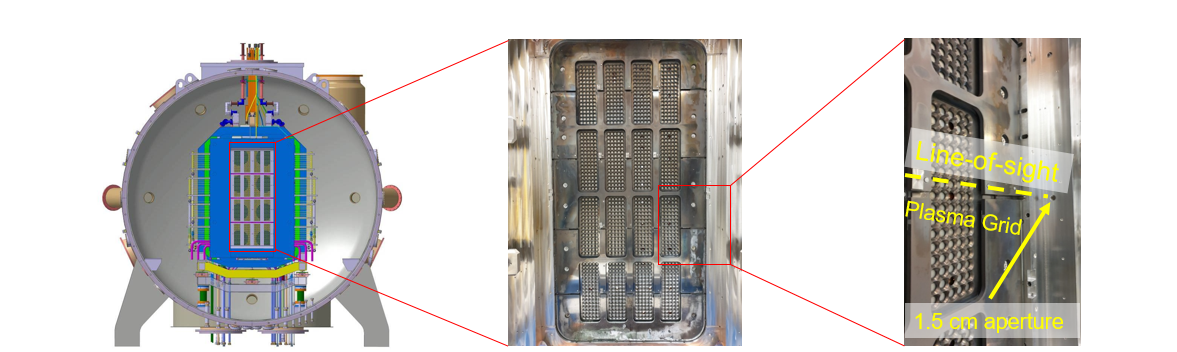}
    \caption{Left: Cross section of the vessel and the source. Center: Photo of the grids facing the plasma. Right: Photo from the interior of the plasma source marking the position of a 1.5 cm aperture at 7 cm from the PG 
}
    \label{fig:SPIDER_section}
\end{figure}

Current negative ion sources for fusion rely on the production of negative hydrogen or deuterium ions ($\mathrm{H^-}$, $\mathrm{D^-}$) by surface assisted processes, enhanced by the presence of caesium deposited onto the plasma-exposed grid, called \textit{plasma grid} (PG) \cite{wunderlich2019,serianni2022}. The negative ions produced in the proximity of PG can however be easily destroyed in the plasma volume before their extraction. The density of negative ions in front of the extraction apertures, in the plasma, is the result of a balance between the dominant surface production processes, in caesium operations, and destruction processes by collisions with electrons, ions, neutral atoms and molecules \cite{bacal2006}. The negative ions destruction processes due to electrons can be partially mitigated by cooling them down by a magnetic field produced by circulating a current through the PG. Although most of the extracted electrons are filtered out by the co-extracted electrons suppression magnets (CESM) embedded in the \textit{extraction grid} (EG), the unfiltered electrons can represent a critical issue during beam extraction operations since they can deposit a significant amount of heat when impinging at high energy on the acceleration grids.  
These issues are currently investigated at the full-scale prototype of a negative ion source for fusion, called SPIDER \cite{toigo2021}, an intermediate experimental step towards the development of the Heating Neutral Beam (HNB) injectors for ITER.
The main technological and scientific goal of SPIDER is maximizing the negative ion current density while maintaining <10\% beam uniformity up to an energy of 100 keV, the highest energy achievable in SPIDER. 
The plasma source is equipped with a broad suite of source and beam diagnostics to drive its optimization towards ITER performance targets \cite{pasqualotto2017}.
To characterize the plasma in the expansion region, both fixed and movable Langmuir probes are employed, together with optical emission spectroscopy based on multiple line-of-sights covering the entire source extension \cite{zaniol2012}.
Although these techniques offer a large overview of plasma source properties, they rely on models that can significantly affect the accuracy of the estimate of electron density and temperature. It would therefore be beneficial to rely on a non-invasive and direct tool such as a diagnostic based on mm-waves.
\newline
In recent years, mm-wave interferometry has been successfully applied in a large negative ion source for fusion \cite{nagaoka2011,tokuzawa2016} but never in a full-size prototype environment, to the knowledge of the authors. The primary difficulties of directly applying this technique in SPIDER are the large dimension of the vessel and the poor accessibility to the plasma source. Fig.\,\ref{fig:SPIDER_section} provides a schematic of the experimental setup, showing the aspect of the PG surface and the location of a candidate aperture (15-mm diameter), currently dedicated to Optical Emission Spectroscopy (OES), through which a probing mm-wave beam might be shine. The distance to the opposite aperture is 1.2 m long and they are located at 7 cm from the PG surface.  
\newline
The aim of this work is to perform a preliminary study of the propagation of a 100 GHz beam provided these design constraints, to assess the feasibility degree of this technique in SPIDER and, possibly, provide an overview of the issues that might be expected when applying this technique to SPIDER-like plasma sources. The choice of the 100 GHz frequency is motivated by the fact of being a compromise between the minimization of beam divergence, the expected range of plasma density and the availability of a 75-110 GHz microwave diagnostic system tested for the first time in the helicon plasma device RAID \cite{furno2017, agnello2020_NF, agnello2020_thomson, simonin2016}.


\section{The mm-wave interferometer}
\begin{figure}
    \centering
    \includegraphics[width=\linewidth]{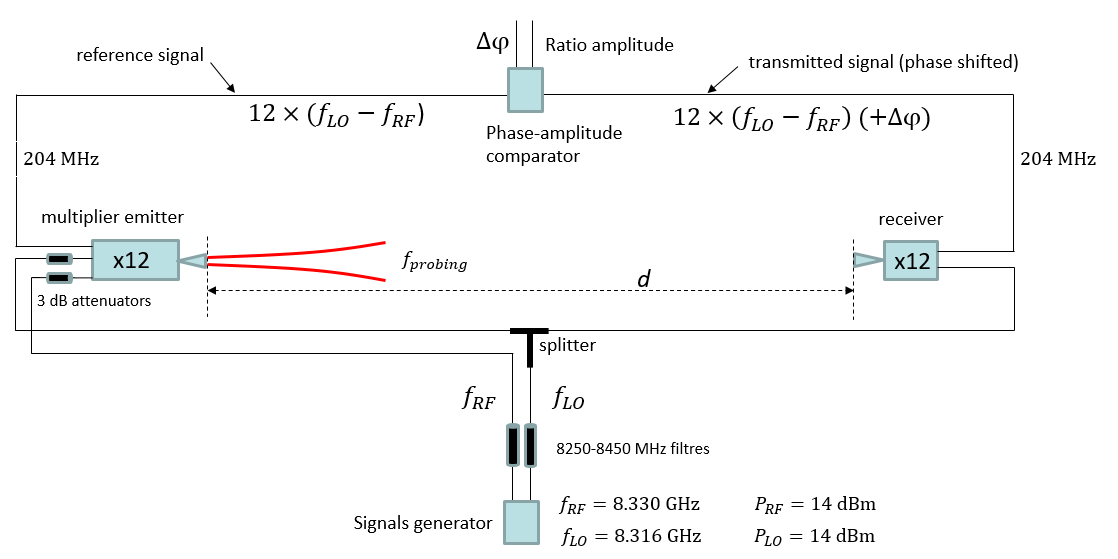}
    \caption{Schematic of the interferometric system}
    \label{fig:scheme_interferometer}
\end{figure}

\begin{figure}
    \centering
    \includegraphics[width=\linewidth]{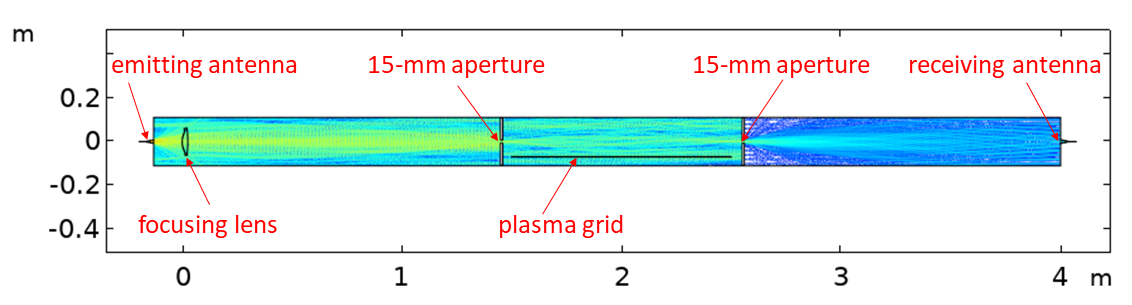}
    \caption{Simulation of a 100 GHz wave propagation in free space across an horizontal section of SPIDER vessel.}
    \label{fig:simulation_wave}
\end{figure}

The interferometer is designed to measure the line-integrated electron density during the plasma discharge with the probing mm-wave beam whose electric field, $\vec{E_1}$, is perpendicular to its propagation direction. The phase difference $\Delta \phi$, compared to vacuum, due to the presence of the electrons, in a unmagnetized plasma, is:
\begin{equation}
    \Delta \phi = \frac{\lambda e^2}{4 \pi c^2 m_e \epsilon_{0}} \int_{LOS} n_e dl;
    \label{eq:simple_phase_shift}
\end{equation}
where $\lambda$ is the wavelength, $e$ the elementary charge, $c$ the speed of light, $m_e$ the electron mass, $\epsilon_0$ the vacuum permittivity and $n_e$ the electron density.
\begin{figure}
    \centering
    \includegraphics[width=8cm]{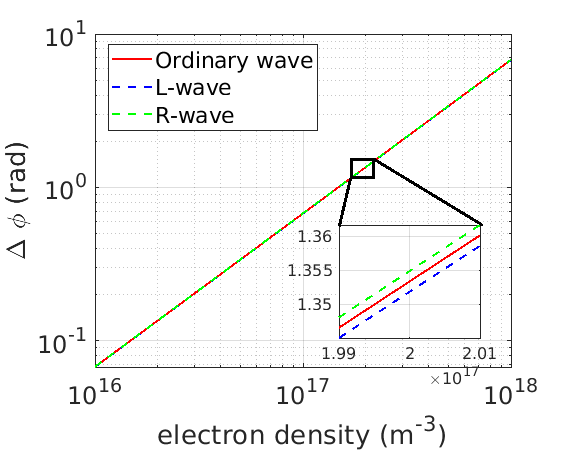}
    \caption{Expected phase shift as a function of the electron density for an effective length of 0.8 m and in the case of an ordinary ($\vec{E_1} \perp \vec{B_0}$) or extraordinary ($\vec{E_1} \parallel \vec{B_0}$) waves and a magnetic field of 4 mT.}
    \label{fig:ordinary_L_R_wave}
\end{figure}
When an external magnetic field is present, as it would be the case if a magnetic filter field is produced in front of the PG, and, the wave propagation direction is parallel to the background magnetic field, the wave dispersion equation is described by two associated waves commonly known as $R$ and $L$ wave. The associated indexes of refraction $N$ of the $R$ and of the $L$ waves are, respectively:
\begin{equation}
    N_{R,L}=\sqrt{1-\frac{\omega_{pe}^2}{\omega(\omega \pm \omega_{ce})}};
\end{equation}
 where $\omega_{pe}$ is the plasma frequency and $\omega_{ce}$ the electron cyclotron frequency. Fig.\,\ref{fig:ordinary_L_R_wave} shows the phase shift with and without the background magnetic field; the difference, at the expected electron density ($2\times 10^{17}\,\mathrm{m}^{-3}$)  is less than 0.5\%. We have therefore used eq.\,\ref{eq:simple_phase_shift} to compute the electron density in this work.
 \newline
We have setup an heterodyne interferometer in the experimental conditions as it would be installed on SPIDER, as shown in the schematic in Fig.\,\ref{fig:scheme_interferometer}.
A signal synthesizer generates a probing reference (RF) and a local oscillator (LO) signal at the frequencies $f_{RF}=8.333\,\mathrm{GHz}$ and $f_{LO}=8.316\,\mathrm{GHz}$, respectively. Both signals are delivered with 14 dBm power and filtered by 200 MHz bandwidth filters. The LO signal is split and sent both to the multiplier emitter and the multiplier receiver, whose frequency multiplication factor is 12, so that the probing frequency is 99.996 GHz. The emitter and the receiver are equipped with circular horn antennae fed by WR-10 standard waveguides. The probing and the reference signals are then down-converted by mixing with the LO frequency at an intermediate frequency (IF) of 204 MHz. The probing and the reference IF signals can be then delivered to the phase-amplitude comparator circuit, whose outputs provide a measure of the phase shift and of the amplitude ratio. We do not have used the phase comparator circuit in our test due to the poor signal-to-noise ratio (S/N).
\newline
As a first required step towards the development of the interferometric technique, numerical simulations of wave propagation were required. Preliminary calculations of a 100\,GHz Gaussian beam envelope in a simplified SPIDER geometry demonstrated that the intensity of the transmitted signal was sufficient to be detected by the receiving module, however, to account for the diffraction effects and to analyse the beam multipaths, full wave 2D numerical simulation in COMSOL\textsuperscript{TM} were performed. Fig.\,\ref{fig:simulation_wave} shows the wave electric field produced by the circular horn antenna, on the left, and focused by a dielectric lens. 
The beam propagates across 1.5 m in free space and it is partially transmitted through the 15-mm aperture. In the space between the two apertures, the wave is reflected by the surrounding metal elements and it is transmitted downstream to the receiver horn antenna. The wave intensity drop from the exit of the emitting horn to the entrance of the receiving horn antenna is approximately 40 dB.
The main outcome of 2D full-wave simulations is that despite the tight wave propagation constraints, a non-negligible amount of power would be transmitted downstream to the receiver, while standing waves would establish inside the source volume. However, this result is not sufficient to assess whether it is possible or not to decouple a phase shift due the one passage of the wave through the investigating medium, such as the plasma, from that due to reflections and multiple-paths. The phase shift is affected by the modules intrinsic phase stability, the mechanical vibrations and the presence of beam reflection multiple-paths.  
It was thus necessary to verify the capability of the interferometric system on a simplified testbench mimicking SPIDER geometry, as described in the next paragraph.
\newline

\section{Experiments on a testbench}

\begin{figure}[htb]
\centering
\begin{subfigure}{0.6\textwidth}
    \includegraphics[width=\textwidth]{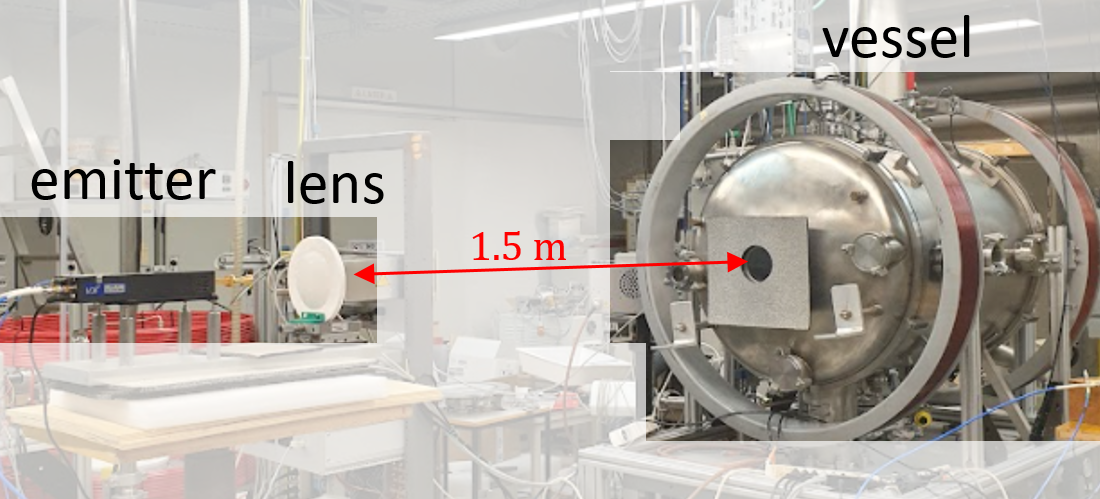}
    \caption{}
    \label{fig:first}
\end{subfigure}
\hfill
\begin{subfigure}{0.35\textwidth}
\includegraphics[width=\textwidth]{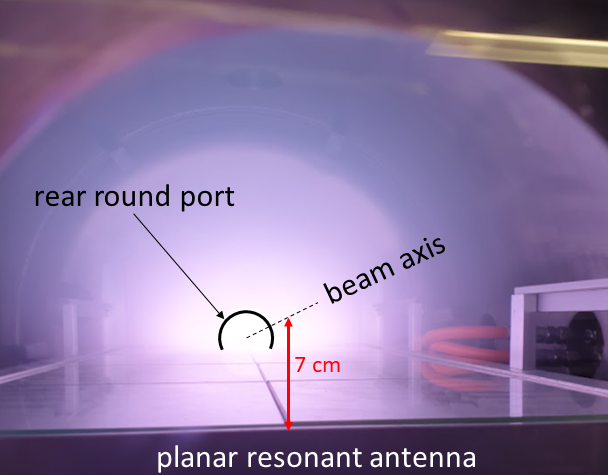}
    \caption{}
    \label{fig:second}
\end{subfigure}
\hfill
 \caption{(a) Test bench mimicking SPIDER geometry including the in-vessel free space propagation across 1.5 m. (b) Visible picture of the source interior during a plasma discharge.}
       \label{fig:setup}
\end{figure}

\begin{figure}[htb]
\centering
\begin{subfigure}{0.45\textwidth}
    \includegraphics[width=\textwidth]{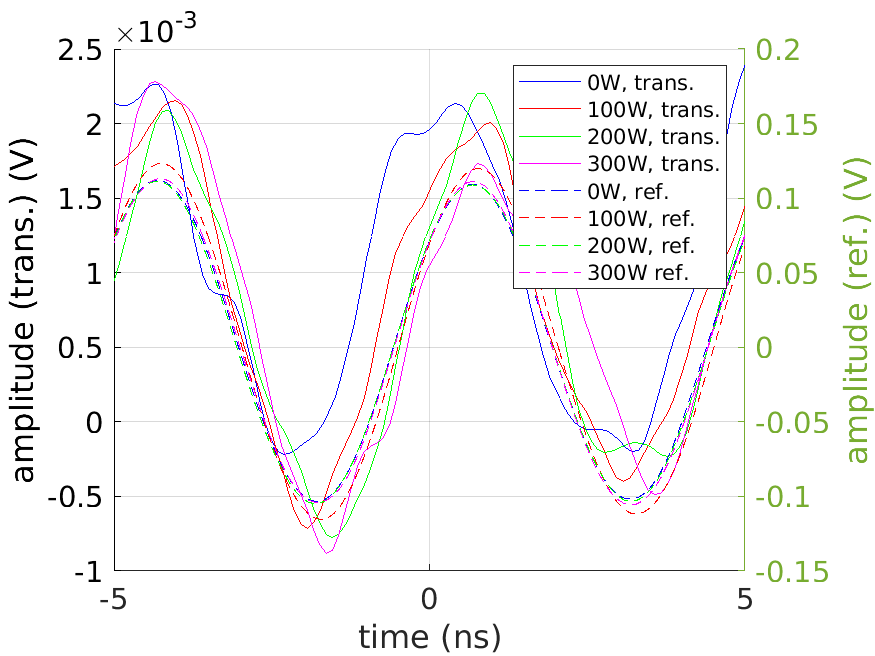}
    \caption{}
    \label{fig:first}
\end{subfigure}
\hfill
\begin{subfigure}{0.45\textwidth}
   \includegraphics[width=\textwidth]{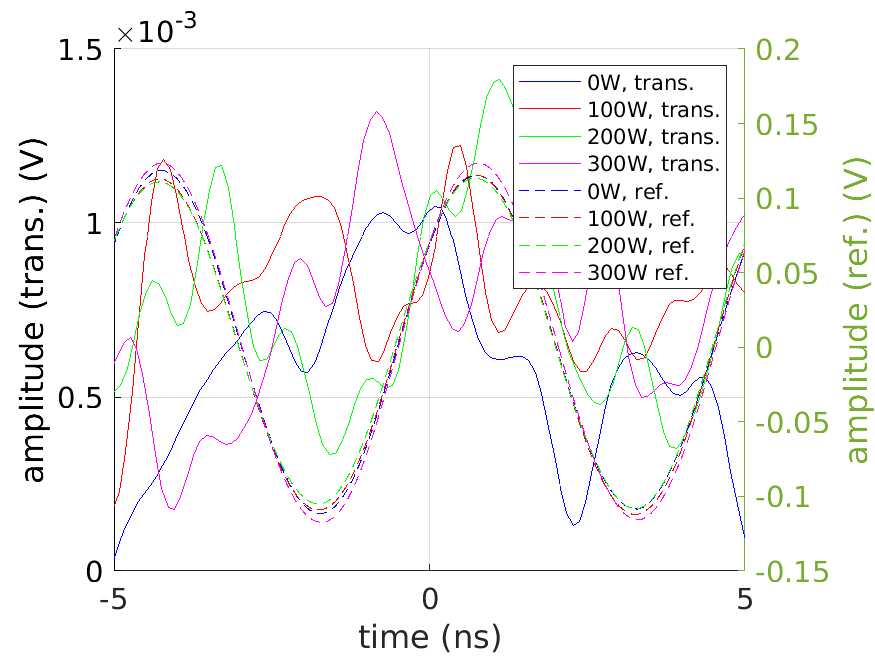}
    \caption{}
    \label{fig:second}
\end{subfigure}
\hfill
\begin{subfigure}{0.45\textwidth}
    \includegraphics[width=\textwidth]{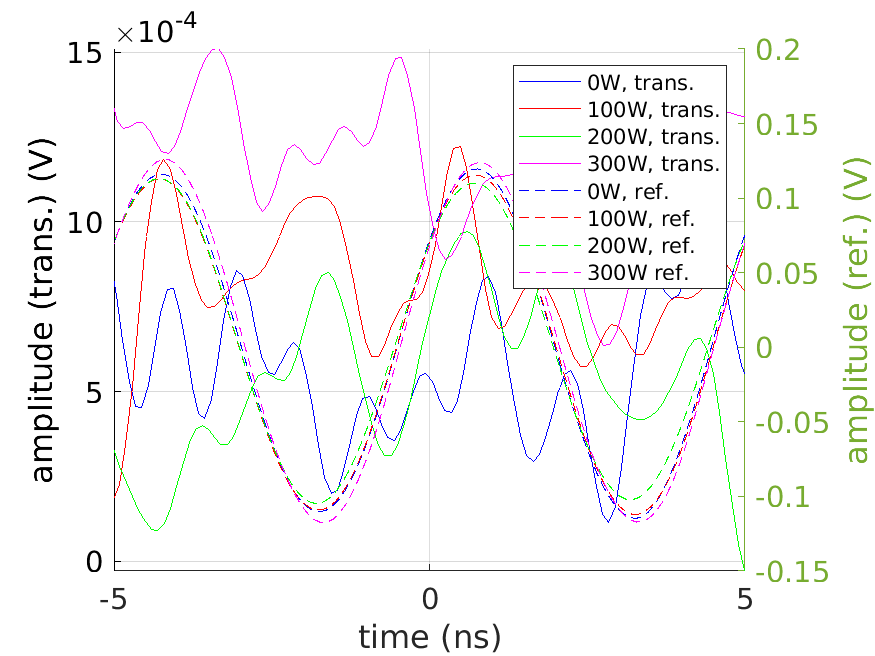}
    \caption{}
    \label{fig:third}
\end{subfigure}
\caption{Mixed probing and reference signals in the three scenarios: (a) with the 80-mm ports, (b) with one 15-mm aperture and (c) with two 15-mm apertures.}
\label{fig:signals_transmitted_experiment}
\end{figure}

\begin{figure}[htb]
\centering
\begin{subfigure}{0.45\textwidth}
    \includegraphics[width=\textwidth]{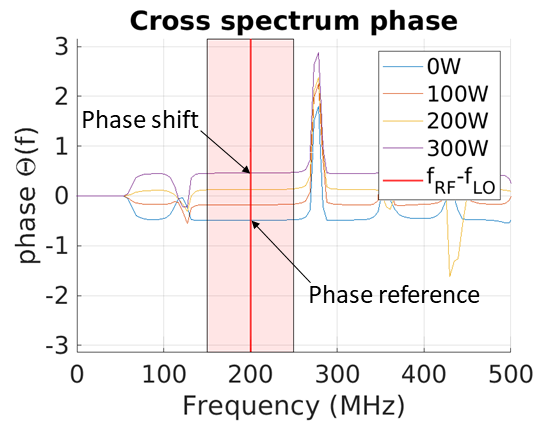}
    \caption{}
    \label{fig:first}
\end{subfigure}
\hfill
\begin{subfigure}{0.45\textwidth}
  \includegraphics[width=\textwidth]{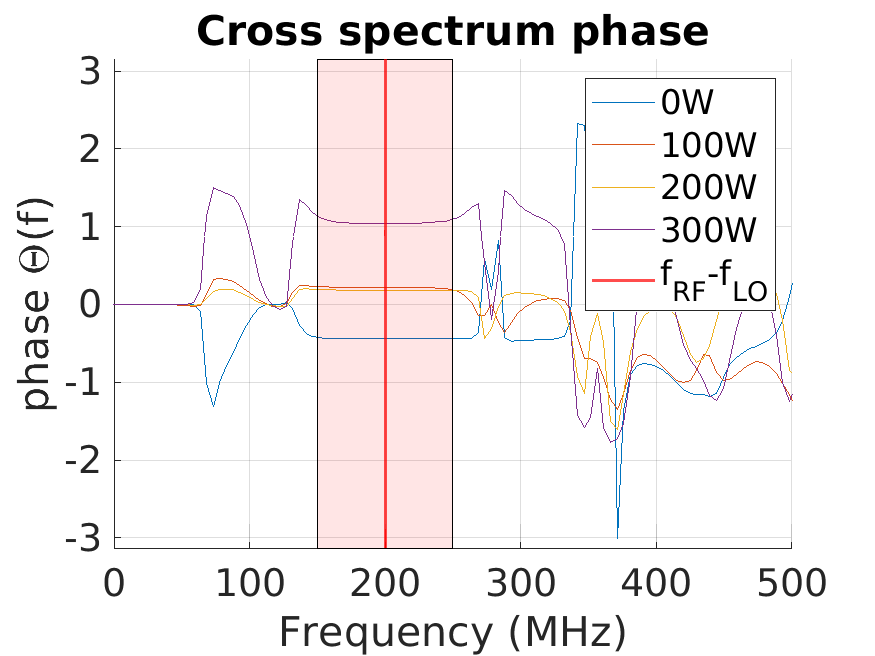}
    \caption{}
    \label{fig:second}
\end{subfigure}
\hfill
\begin{subfigure}{0.45\textwidth}
   \includegraphics[width=\textwidth]{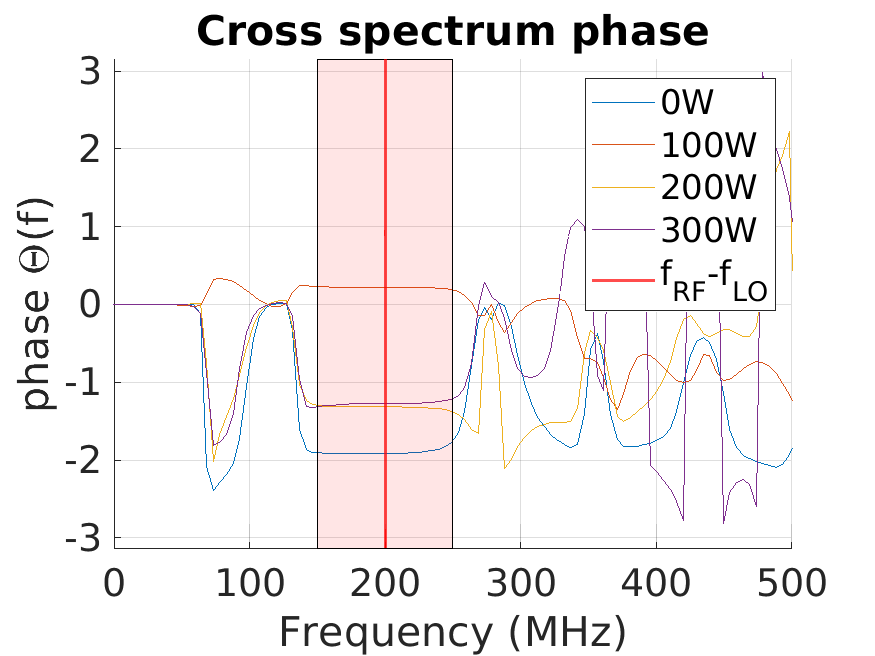}
    \caption{}
    \label{fig:third}
\end{subfigure}
\caption{Phase of the cross power spectral density in the three scenarios: (a) with the 80-mm ports, (b) with one 15-mm aperture and (c) with two 15-mm apertures.}
\label{fig:spectra_phase_cross_correlation}
\end{figure}

The interferometer has been tested in the experimental setup shown in Fig.\,\ref{fig:setup}. It consists of a vacuum vessel in which a plasma can be produced by an inductive planar antenna \cite{guittienne2012} reaching an electron density similar to that measured in the SPIDER expansion region (a few $10^{17}\,\mathrm{m}^{-3}$). 
The vessel is equipped with two axially-aligned 80-mm diameter round ports and the planar antenna dielectric surface is located at 7 cm from the beam propagation axis. The emitter, the focusing lens, the two vessel ports and the receiver are accurately aligned by a visible laser beam and the vessel surface around the 80-mm port is covered by a microwave absorber to minimize back reflections. 
The signal transmitted across the vessel has been measured in three scenarios by increasing beam propagation constraints: with the 80-mm ports and with one or two 15-mm apertures.
Fig.\,\ref{fig:signals_transmitted_experiment} shows the the transmitted (continuous lines) and the reference signals (dashed lines) in the three scenarios, in argon plasma, with a gas pressure of 1 Pa, and increasing values of powers delivered to the plasma discharge: 100 W, 200 W and 300 W. The entire time length of the acquired signals is 20 ms but we only focus in a time range of 10 ns. The time invariance of the reference signals is an evidence of the stability of the apparatus, proving, for instance, the absence of a systematic phase drift over the time to perform the measurements.
Only when the 80-mm ports are used, the oscillations of the transmitted signal follow the trend of the reference signal, albeit the phase shift is not clearly visible in a few oscillations. Once the 15-mm apertures are placed, the S/N drops, thus, making the direct phase comparison unfeasible, therefore, cross correlation analysis with the respective reference signals was performed. 
The cross correlation $\hat{R}_{xy}$ between two discrete time series $x_n$ and $y_n$ is:
\begin{equation}
  \hat{R}_{xy}=\begin{cases}
    \sum_{n=0}^{N-m-1} x_{n+m}y_n^{*}, & m \geq 0.\\
    \hat{R}_{xy}^{*}(-m), & m<0.
  \end{cases}
\end{equation}
The quantity $\hat{R}_{xy}$ measures the overlapping degree of two signals and might result in ambiguities when comparing signals such as those collected in these tests.
To accurately evaluate the frequency dependence of the phase shifts, it is convenient to compute the Cross Power Spectral Density (CPSD), indicated by $P_{xy}$, as follows:
\begin{equation}
    P_{xy}(\omega)=\sum_{m=-\infty}^{\infty} {\hat{R}}_{xy} (m) e^{-j \omega m}.
\end{equation}
By applying numerical filters on CPSD spectra, the cross spectrum phase shift for the three cases are obtained (see Fig.\ref{fig:spectra_phase_cross_correlation}).
Only in the first case the power increase, which is approximately linearly related to the electron density, results in linearly increasing phase shifts. In the region around 200 MHz the phase exhibits an approximate linear increase of approximately  $0.31\pm0.01$ rad, for a 100W power increase.
According to Eq.\,\ref{eq:simple_phase_shift}, at 300 W power this results in an average density of $\simeq 2 \times 10^{17}\,\mathrm{m^{-3}}$ over an effective length of 0.60 m, with an accuracy $\leq 10^{16}\,\mathrm{m^{-3}}$.
However, in the other cases, the position of the phases does not show a linear trend. It is very likely that the internal reflections overcome the one passage signal rendering impossible to isolate the internal multipaths effect.
These results lead to the conclusion that mm-wave interferometry in the current configuration is not feasible and radical modifications of the interferometric apparatus, discussed in the next paragraph, should be carried out.

\section{Proposals for implementation on a full-scale negative ion source}

\begin{figure}[htb]
\centering
\begin{subfigure}{0.37\textwidth}
    \includegraphics[width=\textwidth]{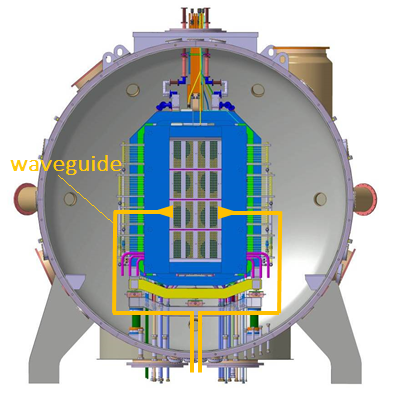}
    \caption{}
    \label{fig:first}
\end{subfigure}
\hfill
\begin{subfigure}{0.48\textwidth}
\includegraphics[width=\textwidth]{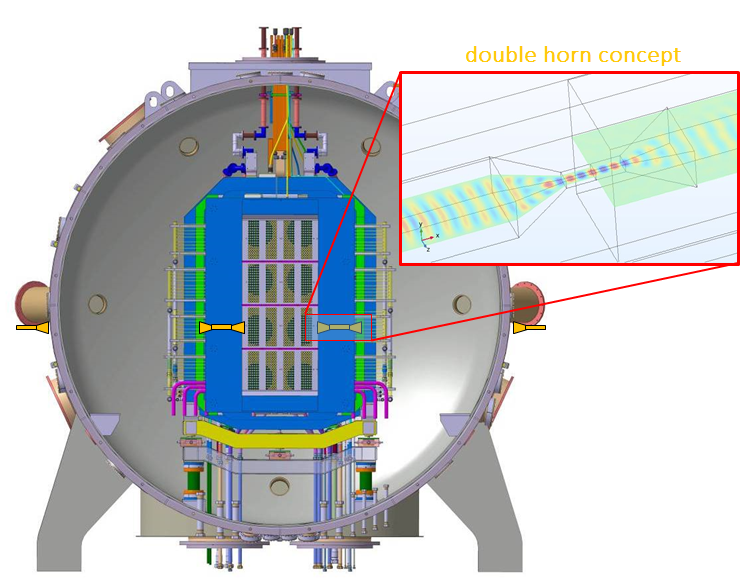}
    \caption{}
    \label{fig:second}
\end{subfigure}
\hfill
 \caption{Proposed schemes to enhance the mm-wave transmission to the plasma source by: (a) full mm-wave channeling; (b) double horn antennae to collect and emit mm-waves.}
\label{fig:mw_transport}
\end{figure}

\begin{figure}
    \centering
    \includegraphics[width=12cm]{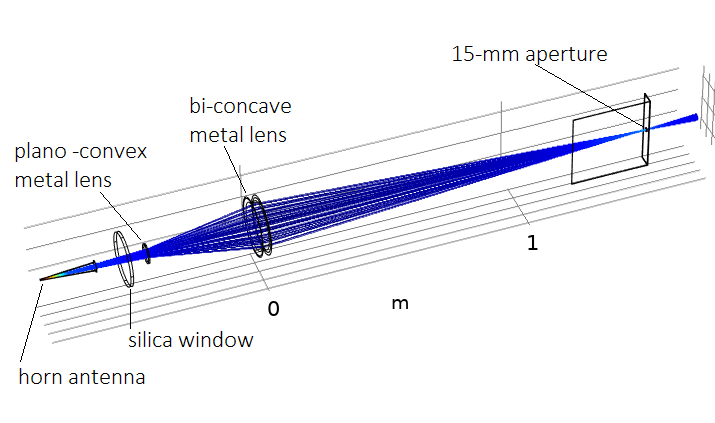}
    \caption{Proposals of an arrangement of metal lenses to focus the mm-wave beam in the 15-mm aperture.}
    \label{fig:metal_lenses}
\end{figure}

Both numerical and experimental investigations reveal that mm-wave interferometry at 100 GHz is impractical on the current SPIDER setup mainly due to the tight spatial constraints on the mm-wave propagation. The strategy to make the technique feasible should be based on the optimization of the transport of the probing mm-wave towards the plasma source.\newline
Two schemes are suggested to enhance the beam transport to the plasma expansion region. Fig.\,\ref{fig:mw_transport}a) shows the case of mm-wave channeling by long waveguide segments; this approach has the advantage of directly shining the beam to the plasma probing region, while avoiding any free wave in the vessel-source volume. Guided transmission losses might represent an issue, however, a first estimate of the power attenuation of the $\mathrm{TE}_{10}$ mode at 100 GHz provides a value of the order of 3 dB/m, therefore power losses are expected to be compatible with the interferometer modules sensitivity. The main technical issue of this scheme is the electrical insulation at the source potential (up to 100 kV) through the electrical bushing, requiring a thorough study of the voltage holding compatibility. To overcome this issue, Fig.\,\ref{fig:mw_transport}b) shows the conceptual scheme of an alternative system requiring only passive in-vessel components. This approach is based on the beam channeling by segments of waveguide segments equipped with horns at their extremities. The idea is that these double horn waveguides could collect and redirect the mm-wave as shown in the inset in Fig.\,\ref{fig:mw_transport}b). 
\newline
Another possible solution to transport mm-waves over large distances could rely on metal lenses, such as in Ref.\cite{tokuzawa2021} where they are employed to shine a 90 GHz beam in the core of a fusion relevant plasma. Compared to dielectric lenses, metallic lenses are plausible candidates in harsh environments.
A metallic lens acts as a medium with a refractive index $n$, less 
 than 1, given by $ n = \bigg[ 1-\bigg(\frac{\lambda}{\lambda_{ co}}\bigg)^2\bigg]^{0.5}$, where $\lambda_{co}$ is the cut-off frequency \cite{goldsmith}. In Fig.\,\ref{fig:metal_lenses} we show, in the ray tracing approximation, the focusing effect of two metal lenses: a plano-convex and a bi-concave, projecting a 100 GHz beam generated by an horn antenna towards the candidate 15-mm aperture of SPIDER source body. Although it might be possible to focus the beam from the injection side, similar focusing elements should be included for its downstream collection.
It would also be worthwhile to consider a variable-frequency probing technique, successfully tested in compact plasma sources such as in Ref.\,\cite{torrisi2016}, where the stray beam reflections can be decoupled from the one passage probing signal.
\newline
As a further step, a specific design of horn antennae would be beneficial to optimize their directivity while maintaining their electrical compatibility with the high voltage environment. For instance, the design of corrugated horns \cite{granet2005} could considerably help in reducing the radiation sidelobes, and, therefore stray beam components that would be reflected on the plasma grid.

\section{Conclusion}
We have discussed a conceptual study of a mm-wave interferometer to perform electron density measurements in the full scale ion source for fusion SPIDER. Numerical simulations and experimental tests on a SPIDER mock up revealed that microwave interferometry at 100GHz might be possible but substantial improvements of the mm-wave transport is essential. The tests on the mockup demonstrated that the sensitivity of the interferometer would in principle allow to perform a phase shift measurements despite the long distance of 4 m at which the modules were located.  The major current constraints is the tight beam accessibility that requires dedicated mm-wave transport by multiple focusing devices. Some solutions have been discussed, however, they would all imply significant interventions on the source, and, the manufacturing of specific components that should be separately tested before their integration in the experiment.

\section{Acknowledgments}
\noindent This work has been carried out within the framework of the EUROfusion Consortium, funded by the European Union via the Euratom Research and Training Programme (Grant Agreement No 101052200 — EUROfusion). Views and opinions expressed are however those of the author(s) only and do not necessarily reflect those of the European Union or the European Commission. Neither the European Union nor the European Commission can be held responsible for them. This work has been carried out within the framework of the ITER-RFX Neutral Beam Testing Facility (NBTF) Agreement and has received funding from the ITER Organization. The views and opinions expressed herein do not necessarily reflect those of the ITER Organization. This work was supported in part by the Swiss National Science Foundation.

\bibliographystyle{unsrt}
\bibliography{BIBLIO}

\end{document}